\documentclass[useAMS,usenatbib]{mn2e}
\usepackage{multirow}
\usepackage{epsfig}
\usepackage{subfigure}
\usepackage{url}

\makeatletter
\let\jnl@style=\rm
\def\ref@jnl#1{{\jnl@style#1}}

\def\aj{\ref@jnl{AJ}}                   
\def\araa{\ref@jnl{ARA\&A}}             
\def\apj{\ref@jnl{ApJ}}                 
\def\apjl{\ref@jnl{ApJ}}                
\def\apjs{\ref@jnl{ApJS}}               
\def\ao{\ref@jnl{Appl.~Opt.}}           
\def\apss{\ref@jnl{Ap\&SS}}             
\def\aap{\ref@jnl{A\&A}}                
\def\aapr{\ref@jnl{A\&A~Rev.}}          
\def\aaps{\ref@jnl{A\&AS}}              
\def\azh{\ref@jnl{AZh}}                 
\def\baas{\ref@jnl{BAAS}}               
\def\jrasc{\ref@jnl{JRASC}}             
\def\memras{\ref@jnl{MmRAS}}            
\def\mnras{\ref@jnl{MNRAS}}             
\def\pra{\ref@jnl{Phys.~Rev.~A}}        
\def\prb{\ref@jnl{Phys.~Rev.~B}}        
\def\prc{\ref@jnl{Phys.~Rev.~C}}        
\def\prd{\ref@jnl{Phys.~Rev.~D}}        
\def\pre{\ref@jnl{Phys.~Rev.~E}}        
\def\prl{\ref@jnl{Phys.~Rev.~Lett.}}    
\def\pasp{\ref@jnl{PASP}}               
\def\pasj{\ref@jnl{PASJ}}               
\def\qjras{\ref@jnl{QJRAS}}             
\def\skytel{\ref@jnl{S\&T}}             
\def\solphys{\ref@jnl{Sol.~Phys.}}      
\def\sovast{\ref@jnl{Soviet~Ast.}}      
\def\ssr{\ref@jnl{Space~Sci.~Rev.}}     
\def\zap{\ref@jnl{ZAp}}                 
\def\nat{\ref@jnl{Nature}}              
\def\iaucirc{\ref@jnl{IAU~Circ.}}       
\def\aplett{\ref@jnl{Astrophys.~Lett.}} 
\def\apspr{\ref@jnl{Astrophys.~Space~Phys.~Res.}}
\def\bain{\ref@jnl{Bull.~Astron.~Inst.~Netherlands}}
\def\fcp{\ref@jnl{Fund.~Cosmic~Phys.}}  
\def\gca{\ref@jnl{Geochim.~Cosmochim.~Acta}}   
\def\grl{\ref@jnl{Geophys.~Res.~Lett.}} 
\def\jcp{\ref@jnl{J.~Chem.~Phys.}}      
\def\jgr{\ref@jnl{J.~Geophys.~Res.}}    
\def\jqsrt{\ref@jnl{J.~Quant.~Spec.~Radiat.~Transf.}}
\def\memsai{\ref@jnl{Mem.~Soc.~Astron.~Italiana}}
\def\nphysa{\ref@jnl{Nucl.~Phys.~A}}   
\def\physrep{\ref@jnl{Phys.~Rep.}}   
\def\physscr{\ref@jnl{Phys.~Scr}}   
\def\planss{\ref@jnl{Planet.~Space~Sci.}}   
\def\procspie{\ref@jnl{Proc.~SPIE}}   

\makeatother

\newcommand {\apgt} {\ {\raise-.5ex\hbox{$\buildrel>\over\sim$}}\ }
\newcommand {\aplt} {\ {\raise-.5ex\hbox{$\buildrel<\over\sim$}}\ } 

\title[Simultaneous {\it NuSTAR} and {\it XMM-Newton} spectroscopy of the NLS1 SWIFT J2127.4+5654]{Simultaneous {\it NuSTAR} and {\it XMM-Newton} 0.5-80 keV spectroscopy of the Narrow Line Seyfert 1 galaxy SWIFT J2127.4+5654}

\author[A. Marinucci, et al.]{A. Marinucci$^{1}$\thanks{E-mail: marinucci@fis.uniroma3.it (AM)},  G. Matt$^{1}$, E. Kara$^{2}$, G. Miniutti$^{3}$, M. Elvis$^{4}$, 
\newauthor
 P. Arevalo$^{5}$, D. R. Ballantyne$^{6}$, M. Balokovi\'{c}$^{7}$, F. Bauer$^{5}$, L. Brenneman$^{4}$,  
\newauthor
 S. E. Boggs$^{8}$, M. Cappi$^{9}$, F. E. Christensen$^{10}$, W. W. Craig$^{10,11}$, A. C. Fabian$^{2}$, 
\newauthor
F. Fuerst$^{7}$, C. J. Hailey$^{12}$,  F. A. Harrison$^{7}$, G. Risaliti$^{4,13}$, C. S. Reynolds$^{14}$, 
\newauthor
D. K. Stern$^{15}$, D. J. Walton$^{7}$ and W. Zhang$^{16}$ \\
$^1$Dipartimento di Fisica, Universit\`a degli Studi Roma Tre, via della Vasca Navale 84, 00146 Roma, Italy\\
$^2$Institute of Astronomy, The University of Cambridge, Madingley Road, Cambridge, CB3 OHA, UK\\
$^3$Centro de Astrobiolog\'ia (CSIC-INTA), Dep. de Astrofisica; ESAC, P.O. Box 78, Villanueva de la Ca\~{n}ada, Madrid, Spain\\
$^{4}$Harvard-Smithsonian Center for Astrophysics, 60 Garden Street, Cambridge, MA, USA\\
$^{5}$Pontificia Universidad Cat\'{o}lica de Chile, Instituto de Astrof\'{i}sica, Casilla 306, Santiago 22, Chile\\
$^{6}$Center for Relativistic Astrophysics, School of Physics, Georgia Institute of Technology, Atlanta, GA 30332, USA\\
$^{7}$Cahill Center for Astronomy and Astrophysics, California Institute of Technology, Pasadena, CA 91125, USA\\
$^{8}$Space Science Laboratory, University of California, Berkeley, California 94720, USA\\
$^{9}$INAF, IASF Bologna, Via P Gobetti 101, 40129 Bologna, Italy\\
$^{10}$DTU Space National Space Institute, Technical University of Denmark, Elektrovej 327, 2800 Lyngby, Denmark\\
$^{11}$Lawrence Livermore National Laboratory, Livermore, California 94550, USA\\
$^{12}$Columbia Astrophysics Laboratory, Columbia University, New York, New York 10027, US\\
$^{13}$INAF Ð Osservatorio Astrofisico di Arcetri, L.go E. Fermi 5, I-50125 Firenze, Italy\\
$^{14}$Department of Astronomy, University of Maryland, College Park, MD 20742-2421, USA\\
$^{15}$Jet Propulsion Laboratory, California Institute of Technology, Pasadena, CA 91109, USA\\
$^{16}$NASA Goddard Space Flight Center, Greenbelt, Maryland 20771, USA\\
}

\begin{document}
\maketitle
\label{firstpage}

\begin{abstract} 
We present a broad band spectral analysis of the joint {\it XMM-Newton} and {\it NuSTAR} observational campaign of the Narrow Line Seyfert 1 SWIFT J2127.4+5654, consisting of 300 ks performed during three {\it XMM-Newton} orbits. We detect a relativistic broadened iron K$\alpha$ line originating from the innermost regions of the accretion disc surrounding the central black hole, from which we infer an intermediate spin of $a$=$0.58^{+0.11}_{-0.17}$. The intrinsic spectrum is steep ($\Gamma=2.08\pm0.01$) as commonly found in Narrow Line Seyfert 1 galaxies, while the cutoff energy (E$_{\rm c}=108^{+11}_{-10}$ keV) falls within the range observed in Broad Line Seyfert 1 Galaxies. 
We measure a low-frequency lag that increases steadily with energy, while at high frequencies, there is a clear lag following the shape of the broad Fe K emission line.  Interestingly, the observed Fe~K lag in SWIFT J2127.4+5654 is not as broad as in other sources that have maximally spinning black holes. The lag amplitude suggests a continuum-to-reprocessor distance of about $ 10-20\ r_{\mathrm{g}}$. These timing results independently support an intermediate black hole spin and a compact corona.
\end{abstract}


\section{ Introduction}
According to the commonly accepted paradigm, luminous Active Galactic Nuclei (AGN) are believed to host a supermassive black hole at their center, surrounded by a geometrically thin accretion disc. The nuclear hard power law continuum that dominates the spectral emission above 2 keV is thought to arise in a hot corona above the accretion disc, where UV/optical seed photons from the disc are Compton scattered towards the X-ray band. This primary X-radiation in turn illuminates the disc and it is partly reflected towards the observer's line of sight \citep[two phase model,][]{hm91,hmg94}. Such a physical process leads to a power-law spectrum extending to energies determined by the electron temperature in the hot corona. The power-law index is a function of the plasma temperature $T$ and optical depth $\tau$. 

The X-ray spectra of unobscured AGN is characterized by ubiquitous and, generally, non-variable features, like the neutral iron K$\alpha$ narrow core and the Compton reflection component \citep{per02, bianchi07}. These features can be attributed to the reprocessing of the nuclear radiation by distant, neutral matter  \citep[i.e. the 'pc-scale torus' in the framework of standard Unification Models,][]{antonucci93}. Furthermore, there are several important spectral features, like the warm absorber, the soft excess, and the relativistic component of the iron K$\alpha$ line, which are present in a number of objects \citep{crummy06, blust05}; different observations of the same object often show these features to be variable \citep{pico05, nan07,delacalle10}. 

X-ray wavelengths offer the best opportunity in which to investigate the physical properties of the primary nuclear source closest to the event horizon and to give constraints on its geometry. Recent works suggest that most of the emission arises from within $\sim20\ r_{\rm g}$ \citep {rm13} and X-ray microlensing experiments confirm that it is compact; in some bright quasars a half-light radius of the corona of $r_{\rm s}\leq$6 $r_{\rm g}$ has been measured \citep{ckd09}, where the gravitational radius is defined as $r_{\rm g}\equiv{ GM/c^{2}}$. Eclipses of the X-ray source have also placed constraints on the size of the hard X-ray emitting region: $r_{\rm s}\leq$30 $r_{\rm g}$ in NGC 1365 \citep[][]{ris07, mrs10, bren13},  $r_{\rm s}\leq$11 $r_{\rm g}$ in SWIFT J2127.4+5654 \citep{sanmi13}. 

Recently, X-ray reverberation around accreting black holes has also revealed that the corona is compact.  On short timescales, the variability associated with the primary continuum is found to lead the variability associated with the soft excess and the broad iron K spectral features.  These time delays are short \citep[e.g. tens of seconds in 1H0707-465, which has a black hole mass of $\sim 2 \times 10^{6}\ M_{\odot}$;][]{kara13a}, suggesting that the light travel time between the continuum emitting corona and the reprocessing region is small, within $\sim 10~r_{\rm g}$.  The first detection of X-ray reverberation was found between the continuum and the soft excess \citep{fabian09}.  In this work, the soft excess was shown to be associated with the broad iron L emission line which peaks at $\sim 0.8$~keV.  Since then, reverberation lags associated with the soft excess and the iron K emission line have been discovered \citep{zfr12,demarco13,kara13c}, and all reveal that the X-ray emission is coming from very small radii.  In this paper, we explore the frequency-dependent time lags in SWIFT J2127.4+5654 that are associated with the soft excess and the broad iron K line.

 \begin{figure}
\begin{center}
\epsfig{file=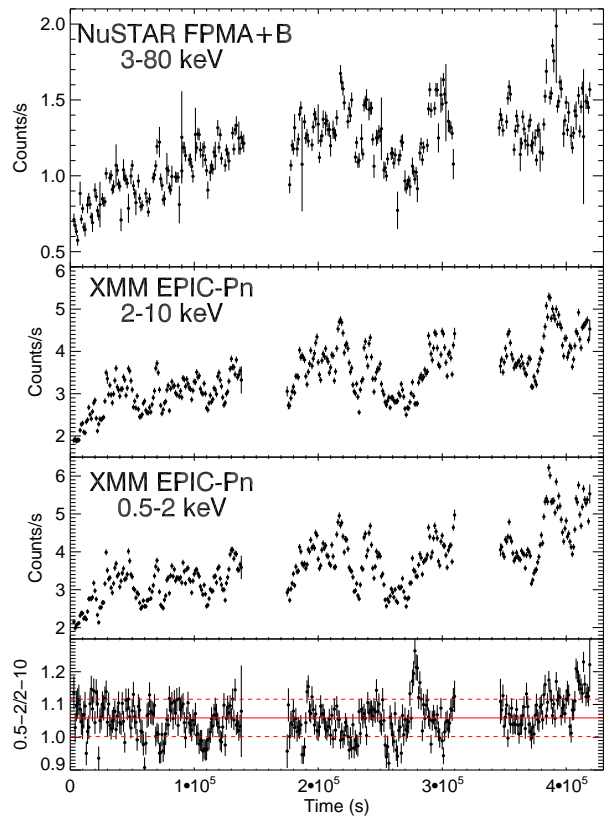, width=\columnwidth}
\caption{\label{lc} From the top to the bottom, {\it NuSTAR} FPMA+B and {\it XMM-Newton} EPIC-Pn light curves in the 3-80 keV, 2-10 keV and 0.5-2 keV energy bands are shown. The ratio between 0.5-2 keV and 2-10 keV EPIC-Pn light curves is shown in the bottom panel, straight and dashed lines indicate mean and standard deviation, respectively.}
\end{center}
\end{figure}
 \begin{figure*}
\begin{center}
\epsfig{file=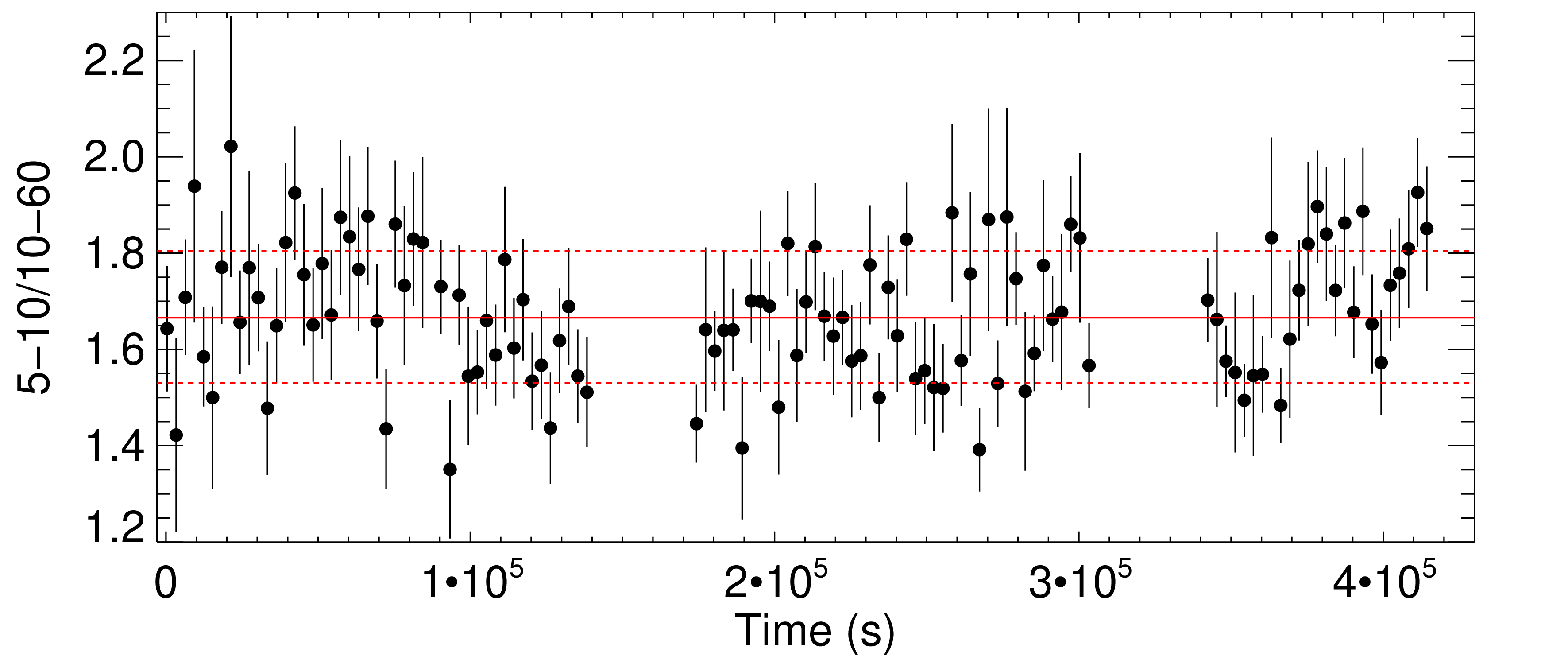, width=\textwidth}
\caption{\label{hr}The ratio between 5-10 keV and 10-60 keV NuSTAR FPMA light curves (in 3000 s bins) is plot versus the time from the start of the NuSTAR pointing, straight and dashed lines indicate mean and standard deviation, respectively.}
\end{center}
\end{figure*}
In addition to these geometric constraints, cutoff energies in several bright Seyfert galaxies have been measured with hard X-ray satellites in the past, such as {\it BeppoSAX} \citep{dad07,per02} and {\it INTEGRAL} \citep{panessa11,derosa12, mbm13}. Current measurements of cutoff energies range between 50 and 300 keV and require spectra extending above 50 keV for better constraints, due to the contribution from complex spectral components (such as reflection from the accretion disc and from distant material) to the broad band spectral shape. 

Narrow Line Seyfert 1 galaxies (NLS1) are a class of AGN whose spectra show peculiar emission-line and continuum properties. They are considered to be Eddington limit accretors \citep{bg92} and to systematically host smaller black hole masses than Broad Line Seyfert 1 galaxies \citep[][and references therein]{kx07}. The primary defining criteria of these objects are the width at half-maximum (FWHM)  of the Balmer emission lines  in their optical spectra ($\aplt 2000$ km s$^{-1}$) and the relative weakness of the [OIII] emission at $\lambda$5007 \citep{op85}.  Past X-ray analyses have shown their very variable continua, on time scales of hours \citep{ppb12}. The X-ray spectra of these objects show very steep slopes, between $\Gamma=2.1-2.5$ \citep{bbf96, bfp96, bme97}.

The {\it Nuclear Spectroscopic Telescope Array} ({\it NuSTAR}: Harrison et al. 2013) is the first high energy focusing X-ray telescope on orbit, $\sim$100 times more sensitive in the 10-80 keV band compared to previous observatories covering these energies, enabling the study of steep spectrum objects at high energies with high precision.

SWIFT J2127.4+5654 (A.K.A. IGR J21277+5656, $z$=0.0144) was classified as a NLS1 galaxy on the basis of the observed FWHM of the H$\alpha$ emission line \citep[$\sim 1180$ km s$^{-1}$:][]{halpern06}.  It has been first detected with {\it Swift/BAT} in the hard X-rays \citep{tueller05}. The source was detected in the hard band with {\it INTEGRAL/IBIS} and an averaged spectrum was obtained by \citet{maba08} in the 17-100 keV band by summing the available on-source exposures (and therefore emission states), resulting in a steep spectrum, in the range $\Gamma=2.4-3$. The authors also discussed a single epoch optical spectrum: the NLS1 classification was confirmed and a black hole mass of $1.5 \times 10^{7} M_{\odot}$ was inferred. The {\it Swift/XRT} data were analyzed by the same authors together with the {\it INTEGRAL/IBIS} averaged spectrum and the presence of a spectral break or of an exponential cutoff in the range of 30--90 keV was suggested \citep{panessa11}.  SWIFT J2127.4+5654 was then observed in 2007 with {\it Suzaku} for a total net exposure of 92 ks, a L$_{\rm Bol}$/L$_{\rm Edd}\simeq 0.18$ and a $\Gamma=2.06\pm0.03$ were measured \citep{mipa09}. A relativistically broadened Fe K$\alpha$ emission line was detected, strongly suggesting that SWIFT J2127.4+5654 is powered by accretion onto a rotating Kerr black hole, with an intermediate spin value of $a$ = $0.6\pm0.2$. This result has been also confirmed by \citet{patrick11} and in a recent work, using a $\sim 130$ ks long {\it XMM-Newton} observation \citep{sanmi13}. The Fe K$\alpha$ shape has been also interpreted in terms of reprocessing in a Compton-thick disc-wind \citep{tatum12}, which would require super-Eddington luminosities \citep{rey12}.

 We present, in the following, results from a simultaneous {\it NuSTAR} and {\it XMM-Newton} observational campaign performed in November 2012. Taking advantage of the unique {\it NuSTAR} energy window, we cover the 0.5--80 keV energy bandwidth. The paper is structured as follows: in Sect. 2 we discuss the joint {\it NuSTAR} and {\it XMM-Newton} observations and data reduction, in Sect. 3 and 4 we present the spectral and lags analyses, respectively. We discuss and summarize the physical implications of our results in Sect. 5 and 6. 
 
\section{Observations and data reduction}
\subsection{NuSTAR}
{\it NuSTAR} (Harrison et al. 2013) observed SWIFT J2127.4+5654 simultaneously with {\it XMM-Newton} with both Focal Plane Module A (FPMA) and B (FPMB) starting on 2012 November 4 for a total of $\sim340$ ks of elapsed time.  The Level 1 data products were processed with the {\it NuSTAR} Data Analysis Software (NuSTARDAS) package (v. 1.1.1). Cleaned event files (level 2 data products) were produced and calibrated using standard filtering criteria with the \textsc{nupipeline} task and the latest calibration files available in the {\it NuSTAR} calibration database (CALDB). Both extraction radii for the source and background spectra are 1.5 arcmin. Custom good time intervals files were used to extract the FPMA and FPMB spectra simultaneously with the three {\it XMM-Newton} orbits.   After this process, the net exposure times for the three observations were 77 ks, 74 ks and 42 ks, for a total Signal-to-Noise Ratio (SNR) of 215, 235 and 186 for the FPMA and 209, 234 and 186 for the FPMB, between 3 and 80 keV. The background-subtracted count rates are $0.631\pm0.003$, $0.798\pm0.004$ and $0.889\pm 0.005$ counts/s for the FPMA, and $0.604\pm0.003$, $0.785\pm0.003$ and $0.876\pm 0.005$ counts/s for the FPMB, between 3 and 80 keV (Fig. \ref{lc}).

No significant variability is observed in the ratio between the 5-10 and 10-60 keV count rates (see Fig. \ref{hr}) during any of the three exposures, suggesting that spectral variability in the {\it NuSTAR} bandpass during these observations is only weak.

The three pairs of {\it NuSTAR} spectra were binned in order to over-sample the instrumental resolution by at least a factor of 2.5 and to have a SNR greater than 3$\sigma$ in each spectral channel. The cross-calibration constant between the two instruments is consistent with 2\% in the three orbits. 

\subsection{XMM-Newton}
SWIFT J2127.4+5654 was observed by {\it XMM-Newton} \citep{xmm} for $\sim$300 ks, starting on 2012 November 4, during three consecutive revolutions (OBSID 0693781701, 0693781801 and 0693781901) with the EPIC CCD cameras, the Pn \citep{struder01} and the two MOS \citep{turner01}, operated in small window and medium filter mode. Data from the MOS detectors are not included in our analysis since they suffered strongly from photon pileup. The extraction radii and the optimal time cuts for flaring particle background were computed with SAS 12 \citep{gabr04} via an iterative process which leads to a maximization of the SNR, similar to the approach described in \citet{pico04}. The resulting optimal extraction radius is 40 arcsec and the background spectra were extracted from source-free circular regions with a radius of about 50 arcsec. 

After this process, the net exposure times were 94 ks, 94 ks and 50 ks for the EPIC-Pn, with 0.5--10 keV count rates of $5.411\pm0.007$, $6.498\pm0.008$ and $7.787\pm 0.013$ counts/s for the three orbits (Fig. \ref{lc}), respectively.  The observation naturally divides into low, medium and high flux states. The 2-10 keV flux of the source during the third orbit (high flux state) is consistent with the one observed by {\it Suzaku} in  2007 \citep{mipa09}.
Spectra were binned in order to over-sample the instrumental resolution by at least a factor of 3 and to have no less than 30 counts in each background-subtracted spectral channel. This allows the applicability of $\chi^2$ statistics. We do not include the 1.8--2.5 keV energy band due to calibration effects associated with the Si/Au edge \citep{smith13}.\\

\section{Spectral analysis}
The adopted cosmological parameters are $H_0=70$ km s$^{-1}$ Mpc$^{-1}$, $\Omega_\Lambda=0.73$ and $\Omega_m=0.27$, i.e. the default ones in \textsc{xspec 12.8.1} \citep{xspec}. Errors correspond to the 90\% confidence level for one interesting parameter ($\Delta\chi^2=2.7$), if not stated otherwise. 

\begin{figure}
\begin{center}
\epsfig{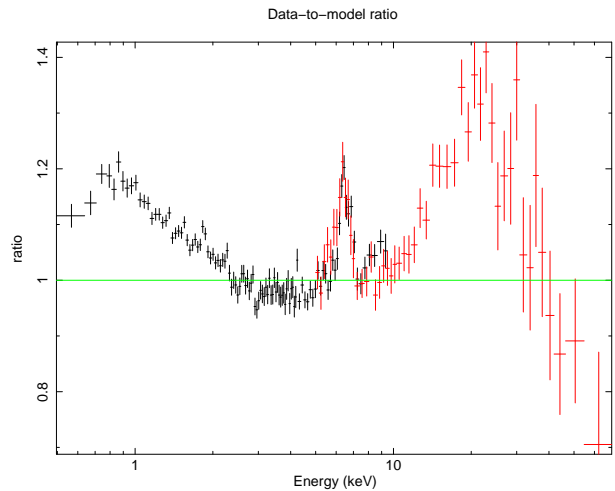}
\caption{\label{ratio} Data-to-model ratio of the medium flux state (Orbit 2). EPIC-Pn and FPMA spectra are shown in black and red, respectively. The continuum is modeled with a  $\Gamma=1.9$ power law (fitted between 2-4 keV and 7.5-10 keV). }
\end{center}
\end{figure}

We started our analysis fitting the medium flux state spectrum with a power law absorbed with only the Galactic column density \citep[$7.65\times10^{21}$ cm$^{-2}$:][]{kalberla05} between 2--4 keV and 7.5--10 keV. In Fig. \ref{ratio} the broad band data-to-model ratio is shown (the continuum is modeled with a  $\Gamma=1.89$ power law) and both a broad iron K$\alpha$ line and a strong Compton hump are clearly visible.

\subsection{\label{EPICanalysis}EPIC PN spectral analysis}
A phenomenological fit is applied in the 3-10 keV range to the three spectra, with a baseline model composed by a power law and four Gaussian lines, to reproduce the broad and narrow components of the Fe K$\alpha$ emission line and the Fe XXV K$\alpha$, Fe XXVI K$\alpha$ lines at 6.7 keV and 6.966 keV, respectively. We left the normalization of the power law free to vary between the three orbits, to include intrinsic flux variations of the source. The fit is good ($\chi^2$/dof=331/321=1.03) and no strong residuals are present; results are presented in Table \ref{ew}. 
Comparing the 2013 observation with the old {\it Suzaku} one, it is interesting noting that the broad component of the iron line ($\sigma=0.22^{+0.23}_{-0.12}$ keV) is less prominent and the Equivalent Width of the narrow Fe K$\alpha$ in the last orbit  (the one comparable in flux with the {\it Suzaku} observation) is consistent with the one reported in \citet{mipa09}.

\begin{table}
\begin{center}
\centering
\hspace{0.3cm}
\begin{tabular}{c|ccccc}
{\bf Id.} & {\bf Energy} &{\bf Flux} & {\bfseries EW$_1$} &{\bfseries EW$_2$} &{\bfseries EW$_3$}\\
 \hline
 \hline
 Fe K$\alpha$ [Br.]& $6.37^{+0.22}_{-0.12}$&$1.3^{+1.4}_{-0.7}$ &$53^{+55}_{-25}$ & $43^{+45}_{-21}$&$38^{+40}_{-18}$ \\
Fe K$\alpha$ [Nar.]&$6.46^{+0.03}_{-0.06}$ & $0.8^{+0.4}_{-0.4}$&$31^{+15}_{-15}$ & $27^{+13}_{-13}$& $23^{+12}_{-12}$\\
  Fe \textsc{xxv} K$\alpha$ & $6.7^*$&$0.5^{+0.3}_{-0.3}$  &$20^{+10}_{-10}$& $17^{+8}_{-8}$&$15^{+7}_{-7}$ \\
   Fe \textsc{xxvi} K$\alpha$ & $6.966^*$& $0.6^{+0.2}_{-0.4}$&$28^{+6}_{-10}$& $24^{+5}_{-8}$&  $21^{+4}_{-7}$\\
\hline
\hline
\end{tabular}\\
\end{center}
\caption{\label{ew}  Best fit parameters for the 3-10 keV {\it XMM} phenomenological fit. Asterisks indicate fixed values, energies are in keV units, fluxes in $10^{-5}$ ph cm$^{-2}$ s$^{-1}$ units and Equivalent Widths for the three orbits are in eV units.}
\end{table}
We then simultaneously fitted the 0.5--10 keV EPIC-Pn spectra from the three orbits with a more physical model, leaving the normalizations of the primary power law and of the disc reflection component free to vary, to take into account the flux variations of the source. When we use a model composed of an absorbed power law and cold distant reflection  the fit is not good ($\chi^2$/dof=849/457=1.86) and strong residuals around the iron K$\alpha$ band are present.

\begin{figure} 
\begin{center}
\epsfig{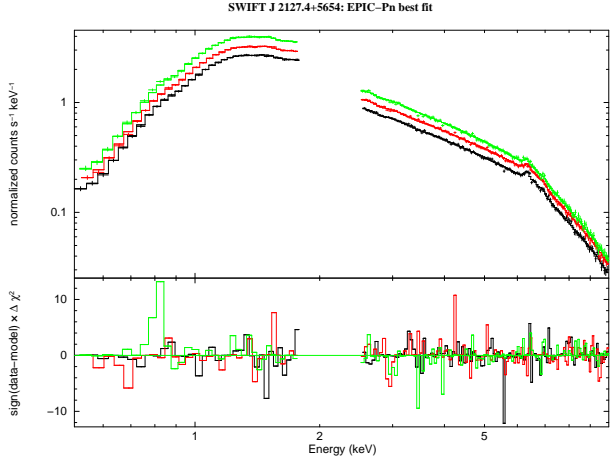}
\caption{\label{bestfitxmm} EPIC-Pn 0.5-10 keV best fit and residuals with a model where blurred relativistic reflection is taken into account, no structured residuals are seen. }
\end{center}
\end{figure}

We then added a relativistically blurred reflection component arising from an ionized accretion disc and a further absorber at the distance of the source \citep{sanmi13}. 
We used \textsc{xillver} for both cold and ionized reflection \citep{garcia13} and \textsc{relconv} for relativistic smearing \citep{dauser13}. The inclusion of a blurred reflection component in the model greatly improves the fit ($\chi^2$/dof=507/451=1.12). The addition of two emission lines at 6.7 keV (Fe \textsc{xxv} K$\alpha$, actually a triplet) and at 6.966 keV  (Fe \textsc{xxvi} K$\alpha$), with fluxes consistent with the ones reported in Table \ref{ew}, marginally improves the fit $\chi^2$/dof=498/449=1.11. Replacing the intrinsic neutral absorber with an ionized one we find an upper limit to the ionization state of $\log \xi \leq -0.54$, leading to no improvement of the fit. The photon index of the primary continuum is steep ($\Gamma=2.11\pm0.01$), as commonly found for NLS1 galaxies and no strong residuals are present throughout the energy band, with the exception of a feature in the spectrum from the third orbit around $\sim0.8$ keV which is not attributable to any known spectral feature (Fig. \ref{bestfitxmm}). The excess in the soft energy band, shown in Fig. \ref{ratio}, is a combination of the steep primary continuum and the blurred ionized reflection, no additional components to the model described above are needed.  We find an intermediate value for the black hole spin ($a$=$0.5\pm0.2$),  in good agreement with the value measured by Suzaku, values for the emissivity and inclination of the disc are also consistent with previous analyses \citep{mipa09}. Best fit parameters are presented in Table \ref{refl_best_fit}. When we leave the column density of the neutral local absorber and the photon indices of the primary continua free to vary between the three spectra a marginal improvement of the fit ($\chi^2$/dof=481/445=1.07) is found, with no variations from combined best fit parameters.

\begin{figure*}
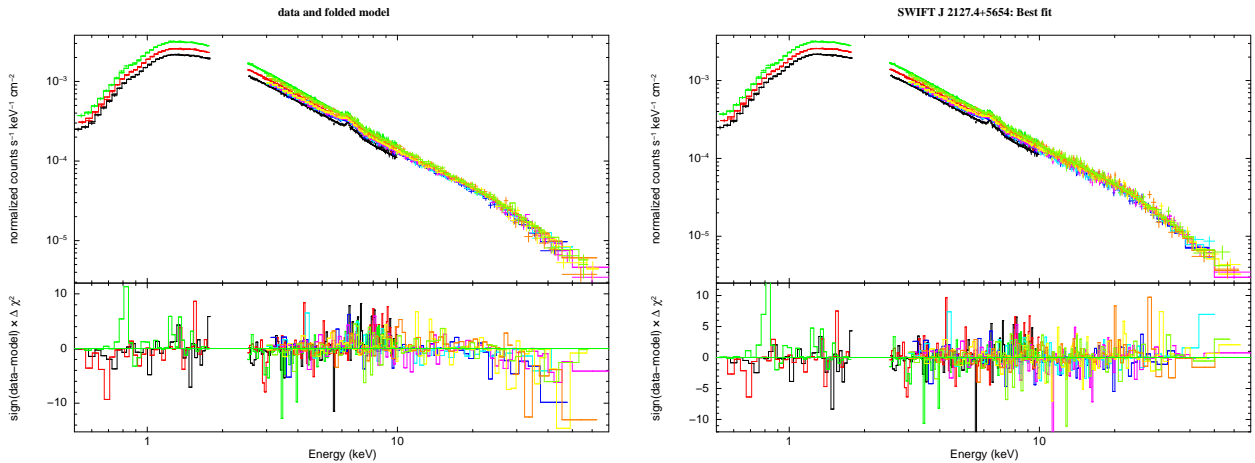
 
\begin{center}
\epsfig{file=bad_combinedfit.eps, angle=-90, width=\columnwidth}
\epsfig{file=best_fit.eps, angle=-90, width=\columnwidth}
\caption{\label{badcombined} \textit{Left panel:} broad band fit of the {\it NuSTAR} and {\it XMM} spectra with a model that does not take into account the high energy cutoff. Clear residuals can be seen above $\sim 20$ keV. \textit{Right panel:} broad band {\it XMM} +{\it NuSTAR} best fit when a cutoff energy is properly modelled.}
\end{center}
\end{figure*}

\begin{table}
\begin{center}
\hspace{0.3cm}
\begin{tabular}{c|cc}
& {\bfseries XMM} &{\bfseries XMM+NuSTAR} \\
 \hline
 \hline
  N$_{\rm H}$ & $2.2\pm0.1$&$2.13\pm0.05$ \\
  $\Gamma$ & $2.11\pm0.01$& $2.08^{+0.01}_{-0.01}$\\
    E$\rm _c$ & -& $108^{+11}_{-10}$\\
  N$_{\rm pow}^1$ ($\times 10^{-3}$) & $9.6 \pm0.1$&$9.6 \pm0.1$ \\
   N$_{\rm pow}^2$ & $11.4 \pm0.1$&$11.4 \pm$ 0.1\\
    N$_{\rm pow}^3$& $14.1 \pm0.1 $&$13.9 \pm$ 0.1\\
  \hline
  N$_{\rm neutral}$ ($\times 10^{-4}$)& $1.25\pm0.25$&$0.95\pm0.15$ \\
  \hline
  q & $6.1^{+2.1}_{-1.7}$&$6.3^{+1.1}_{-1.0}$ \\
  a & $0.5\pm0.2$&$0.58^{+0.11}_{-0.17}$\\
  i & $48^{+5}_{-3}$ & $49\pm2$\\
  $\xi$ & $<20$ &$<8$\\
  $A_{\rm Fe}$ & $0.54^{+0.06}_{-0.04p}$ &$0.71^{+0.05}_{-0.05}$\\
   N$_{\rm refl}^1$ ($\times 10^{-4}$)&$2.9\pm0.5$  &$2.7\pm0.3$\\
   N$_{\rm refl}^2$ &$3.7\pm0.5$  &$3.5\pm0.3$\\
   N$_{\rm refl}^3$ & $3.0\pm0.5$ &$3.0\pm0.4$ \\
        \hline
   $\chi^2$/dof & 498/449=1.11 &1733/1566=1.10\\

\hline
\hline
\end{tabular}\\
\end{center}
\caption{\label{refl_best_fit} Best fit parameters for reflection and absorption models. Column densities are in $10^{21}$ cm$^{-2}$ units, disc inclination angles are in degrees, ionization parameters $\xi$ are in erg cm s$^{-1}$ units, Fe abundance is with respect to the solar value and cutoff energy E$_{\rm c}$ is in keV units. Power law normalizations are in  ph cm$^{-2}$ s$^{-1}$ keV$^{-1}$ at 1 keV and $q$ is the adimensional parameter for the disc emissivity. The subscript $p$ indicates that the parameter pegged at its minimal allowed value.}
\end{table}

\begin{table}
\begin{center}
\centering
\hspace{0.3cm}
\begin{tabular}{c|ccc}
& {\bfseries Orb. 1} &{\bfseries Orb. 2} &{\bfseries Orb. 3}\\
 \hline
 \hline
  F$_{\rm 2-10\ keV}$ &$2.40\pm0.02$ & $2.87\pm0.02$&$3.32\pm0.02$ \\
  L$_{\rm 2-10\ keV}$ & $1.13\pm0.01$&$1.35\pm0.05$ & $1.57\pm0.05$\\
 L$_{\rm 3-80\ keV}^{\rm pow}$ & $1.61\pm0.02$&$1.92\pm0.02$ & $2.35\pm0.02$\\
   F$_{\rm 20-100\ keV}$ & $2.67\pm0.03$&$3.21 \pm0.03$&$3.34\pm0.03$ \\
     L$_{\rm 20-100\ keV}$ &$1.25\pm0.01$&$1.50\pm0.05$ & $1.57\pm0.05$\\
   L$_{\rm bol}$/L$_{\rm Edd}$ & $\simeq0.11$&$\simeq0.14$ & $\simeq0.16$\\
\hline
\hline
\end{tabular}\\
\end{center}
\caption{\label{fluxes} Best fit fluxes and luminosities in different energy intervals. Fluxes are in 10$^{-11}$ erg cm$^{-2}$ s$^{-1}$ units and are observed. Luminosities are in 10$^{43}$ erg s$^{-1}$ units and are corrected for absorption. We calculated L$_{\rm bol}$ from the 2-10 keV luminosity, applying the relation of \citet{mar04} for the bolometric corrections \citep[this factor is also consistent with the lowest bolometric correction for this accretion rate presented in][]{vf09}. A value for the black hole mass of 1.5$\times10^7$ M$_{\odot}$ has been used \citep{maba08}.}
\end{table}

\subsection{\label{nustaranalysis}Broad band spectral analysis}
We then included the six 3--80 keV {\it NuSTAR} (FPMA and FPMB) spectra in our analysis, extending the spectral coverage up to 80 keV. We left the cross-calibration constants between {\it XMM} and {\it NuSTAR} free to vary. A  photon index of $\Gamma=2.08\pm0.01$ is found, in agreement with previous {\it Suzaku} observations \citep{mipa09}. We get a best fit $\chi^2$/dof=1907/1567=1.21 and strong, structured residuals can be seen above $\sim 20$ keV (Fig. \ref{badcombined}, left panel), suggesting the presence of a high energy cutoff. 

\begin{figure*}
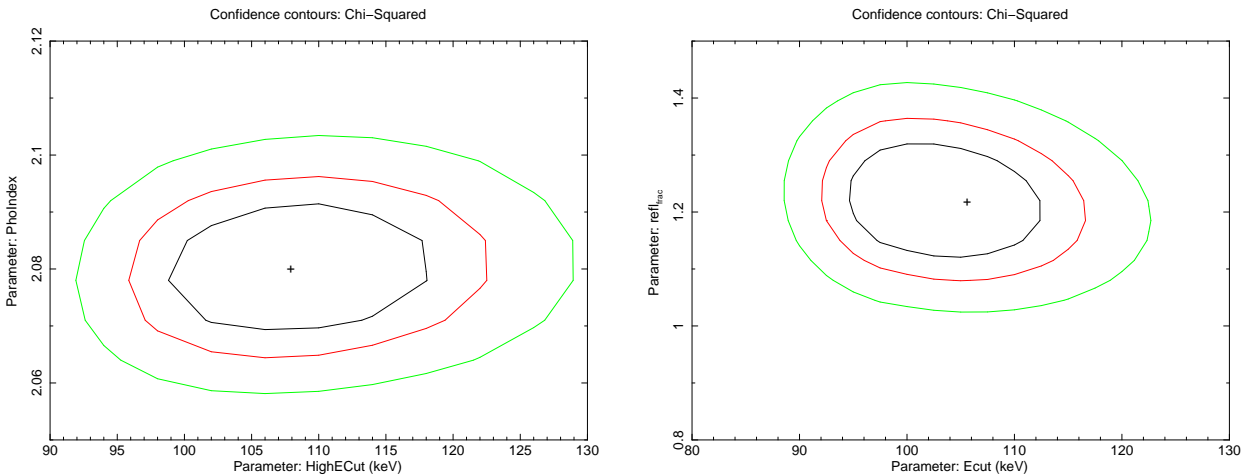
 
\begin{center}
\epsfig{file=cutoff.eps, angle=-90, width=\columnwidth}
\epsfig{file=cont_cut_R.eps, angle=-90, width=\columnwidth}
\caption{\label{contour}  $\Delta\chi^2=$2.30, 6.17 and 11.83 contours (corresponding approximately to 68\%, 95\% and 99.7\% confidence levels) for the cutoff energy E$_{\rm c}$ and photon index $\Gamma$ (left panel) and for the cutoff energy and the blurred ionized reflection fraction in the second orbit (right panel).}
\end{center}
\end{figure*}

Therefore, we substituted the primary power law with the \textsc{cutoffpl} model in \textsc{Xspec} and tied the values of the cutoff energies to the ones of the reflection components.\footnote[1]{A test version of \textsc{xillver} has been used, where the cutoff energy is a variable parameter.} The fit greatly improves ($\Delta \chi^2=174$, for one additional degree of freedom), a $\chi^2$/dof=1733/1566=1.11 is found and no strong residuals are present in the whole energy band (Fig. \ref{badcombined}, right panel). Best fit parameters can be found in Table \ref{refl_best_fit} and fluxes, luminosities and Eddington ratios in Table \ref{fluxes}. The cross-calibration factors between the Pn and the FPMA, FPMB detectors are $\rm K_{\rm FPMA}=1.045\pm0.007$ and $\rm K_{\rm FPMB}=1.074\pm0.007$. We measure a cutoff energy E$\rm _c$=$108^{+11}_{-10}$ keV and $\Gamma=2.08\pm0.01$: Fig. \ref{contour} (left panel) presents the contour plot of photon index versus cutoff energy. 
In order to estimate the relative strength of the Compton hump (arising from both distant and blurred ionized reflection components) with respect to the primary continuum, we followed a method similar to the one described in Walton et al. (2013) for NGC 1365, using the relative normalizations of the reflection (modeled with \textsc{pexrav}, with the inclination angle and iron abundance fixed to best fit values) and of the power law components, above 10 keV. We find values of $R_1=1.5\pm0.2$, $R_2=1.6\pm0.2$ and $R_3=1.1\pm0.2$ for the three sets of spectra, respectively.

 Thanks to the {\it NuSTAR}+{\it XMM} broad band view, the degeneracy between the contribution from the ionized and the cold reflector can be broken, using recent X-ray disc reflection models that include relativistic blurring and a direct measure of $R$. We substituted the blurred ionized reflection and primary components in the broad band best fit model for \textsc{RELXILL} \citep[a model that includes both relativistic effects and reflection from an accretion disc,][]{gdl13}. The reflection fraction $R$ has been left to vary between the three orbits. The overall fit is good ($\chi^2$/dof=1756/1570=1.11) and no differences from best fit parameters presented in Table \ref{refl_best_fit} are found. The contribution of the blurred reflection component from the disc to the total reflection fraction in the three orbits is  $R_1^{disc}=1.1\pm0.1$, $R_2^{disc}=1.2\pm0.1$ and $R_3^{disc}=0.9\pm0.1$. The contour plot of the cutoff energy E$\rm _c$ and the $R_2^{disc}$ parameter is shown in Fig. \ref{contour}, right panel.  If we consider a lamp-post geometry and use \textsc{RELXILLLP} (with spin and inclination parameters fixed to their best-fit values) we find an height of the X-ray source of $25\pm10\ r_g$, consistent with the estimates discussed in Sect. 5.

\begin{figure} 
\begin{center}
\epsfig{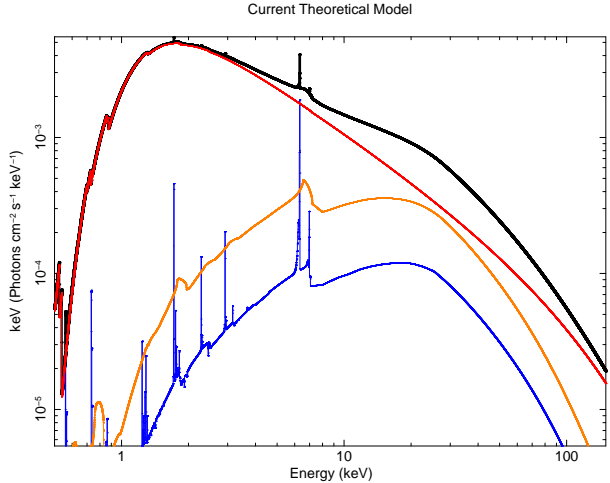}
\caption{\label{model} Relativistic reflection model between 0.5 and 150 keV. Different components such as the relativistic blurred ionized reflection (in orange), the primary continuum (in red) and neutral reflection from distant material (in blue) can be clearly seen.}
\end{center}
\end{figure}

Once a best fit is obtained we try to reproduce the primary continuum (modeled with \textsc{cutoffpl} in \textsc{Xspec} so far) with a more physical model where the electron energy (kT$_e$) and the coronal optical depth ($\tau$) can be disentangled, using different geometries. Photon index and cutoff energy are fixed to their best fit values in the reflection components. We use the \textsc{compTT} model in \textsc{Xspec} \citep{tit94}, which assumes that the corona is distributed either in a slab or a spherical geometry. In such a model the soft photon input spectrum is a Wien law; we fixed the temperature to 50 eV, appropriate for M$_{\rm BH}\approx 10^7$M$_{\odot}$. In the case of a slab geometry the fit is good ($\chi^2$/dof=1733/1569=1.10) and best fit values of kT$_e$=$68^{+37}_{-32}$ keV and $\tau=0.35^{+0.35}_{-0.19}$ are found. When a spherical corona is considered we find a statistically equivalent fit ($\chi^2$/dof=1734/1569=1.10) and best fit values are kT$_e$=$53_{-26}^{+28}$ keV and $\tau=1.35_{-0.67}^{+1.03}$. In both geometries no significant variations from the best fit values in Table \ref{refl_best_fit} are found and 3-80 keV fluxes of the primary continuum are consistent with the ones presented in Table \ref{fluxes}. As already pointed out in \citet{bmf14} in the case of IC 4329A, the difference in optical depth when two different geometries are taken into account is primarily due to the different meaning of this parameter in the two geometries: the optical depth for a slab geometry is taken vertically, while that for a sphere is taken radially \citep[see][for a more detailed description of the models]{tit94}.

\section{\label{lags} XMM-Newton lags}

\begin{figure*} 
\begin{center}
\epsfig{file=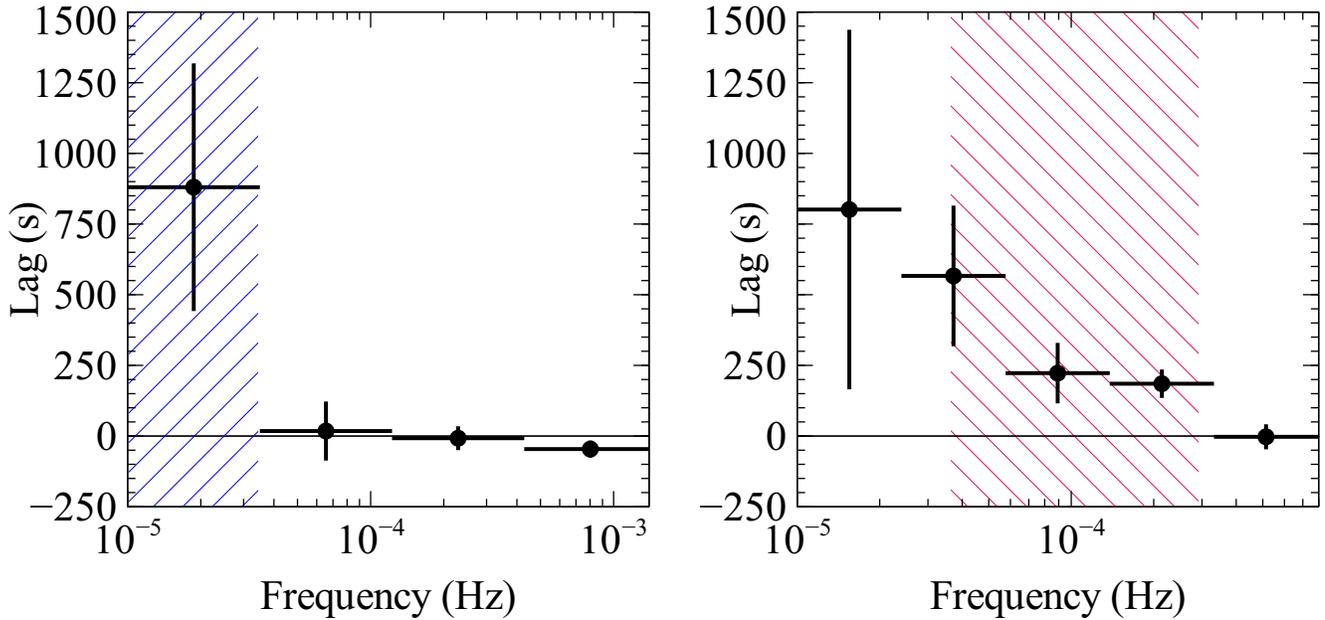, width=\textwidth}
\caption{\label{lagfreq} \textit{Left panel:} The frequency-dependent lag between 0.3--1~keV and 1--5~keV. A positive lag indicates that the hard band lags behind the soft band.  We see a hard lag at $\nu < 3.5 \times 10^{-5}$~Hz, and zero lag above that frequency. Poisson noise begins to dominate the power spectrum at $\nu \sim 5 \times 10^{-4}$~Hz.  (The blue hash shows the frequency range explored for the lag-energy spectrum in the left panel of Fig.~\ref{lagen}).  \textit{Right panel:} The frequency-dependent lag between the 3--5~keV and 5--8~keV bands.  The hard lag extends up to $3\times 10^{-4}$~Hz. (The red hash shows the frequency range for the lag-energy spectrum in the right panel of Fig.~\ref{lagen}).}
\end{center}
\end{figure*}

\begin{figure*} 
\begin{center}
\epsfig{file=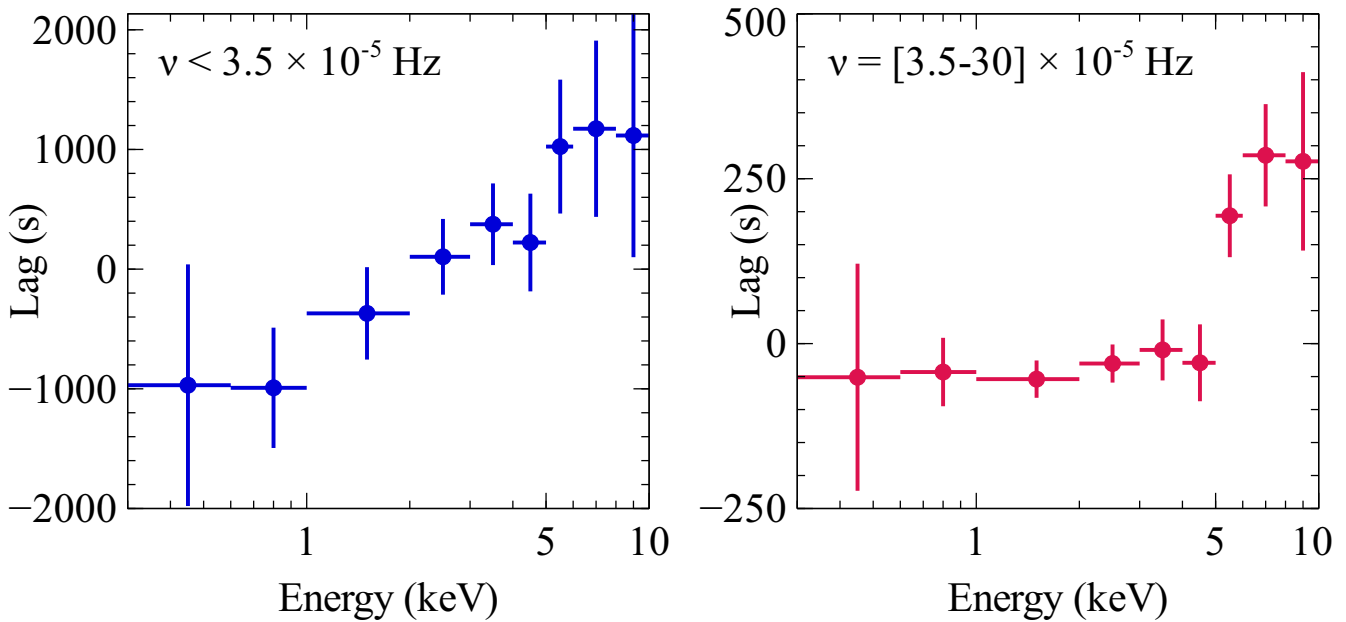, width=\textwidth}
\caption{\label{lagen} \textit{Left panel:} The lag-energy spectrum for low frequencies, $\nu <3.5 \times 10^{-5}$~Hz (corresponding to the frequency range shown in blue hash in Fig.~\ref{lagfreq}. The lag increases steadily with energy.  \textit{Right panel:} The lag-energy spectrum for frequencies $\nu=[3.5-30]10^{-5}$~Hz (the red hash in the right panel of Fig.~\ref{lagfreq}.  This shows little lag between the continuum and soft excess, and a sharp increase in the lag at $>5$~keV.}
\end{center}
\end{figure*}

Finally, in addition to this broad-band spectral analysis, we looked for time lags using the {\em XMM-Newton} observations.  The lags were computed using the standard Fourier technique, described in \citet{nowak99}, where the phase lag is found between the Fourier transform of two light curves.  The phase lag is converted into a frequency-dependent time lag by dividing by $2\pi f$, where $f$ is the temporal frequency.  

We start in usual way by looking for lags associated with the soft excess \citep[as in][]{fabian09,demarco13}.  
The left panel of Fig.~\ref{lagfreq} shows the frequency-dependent lag between 0.3--1~keV and 1--5~keV.  A positive lag indicates a time delay in the hard band, which we find at frequencies below $3.5\times 10^{-5}$~Hz.  The lag goes to zero above this frequency.  We then probe the frequency-dependent lag between 3--5~keV and 5--8~keV, where there is less obscuration (right panel of Fig.~\ref{lagfreq}). This shows the hardi 5--8~keV band lagging at higher frequencies, as well (at $\nu < 30 \times 10^{-5}$~Hz).    

To investigate further, we compute the `lag-energy spectrum', to find the average time delays associated with the energy spectrum.  The zero-point of the lag is dictated by the chosen reference band (for this analysis, it was the entire 0.3--10~keV, excluding the band of interest), and so it is the relative lag between energy bins that is meaningful. The left panel of Fig.~\ref{lagen} shows the lag-energy spectrum for frequencies less than $3.5 \times 10^{-5}$~Hz (where we found the 1--5~keV band lagging behind the 0.3--1~keV band).  The time delay increases steadily with energy, without any obvious spectral features, similar to the low-frequency lag-energy spectra in several other AGN \citep{kara13b,walton13}.  We then probe higher frequencies, from $\nu=[3.5-30] \times 10^{-5}~Hz$, where the 5--8~keV band lags the 3--5~keV band (right panel of Fig.~\ref{lagen}).  The high-frequency lag-energy spectrum shows little to no lag between the soft excess and the 1--5~keV band, but then jumps up 200~s at $\sim$5~keV.  The lag peaks at 6--8~keV, and the points from 5--8~keV are $> 4 \sigma$ above the continuum lag.  It does not show the obvious decrease above 8~keV that is often seen in high-frequency lag-energy spectra, but statistics are low at this high-frequency, so we cannot rule out a downturn of the lag at the highest energies.  

\section{\label{discussion} Discussion}
\begin{figure} 
\begin{center}
\epsfig{file=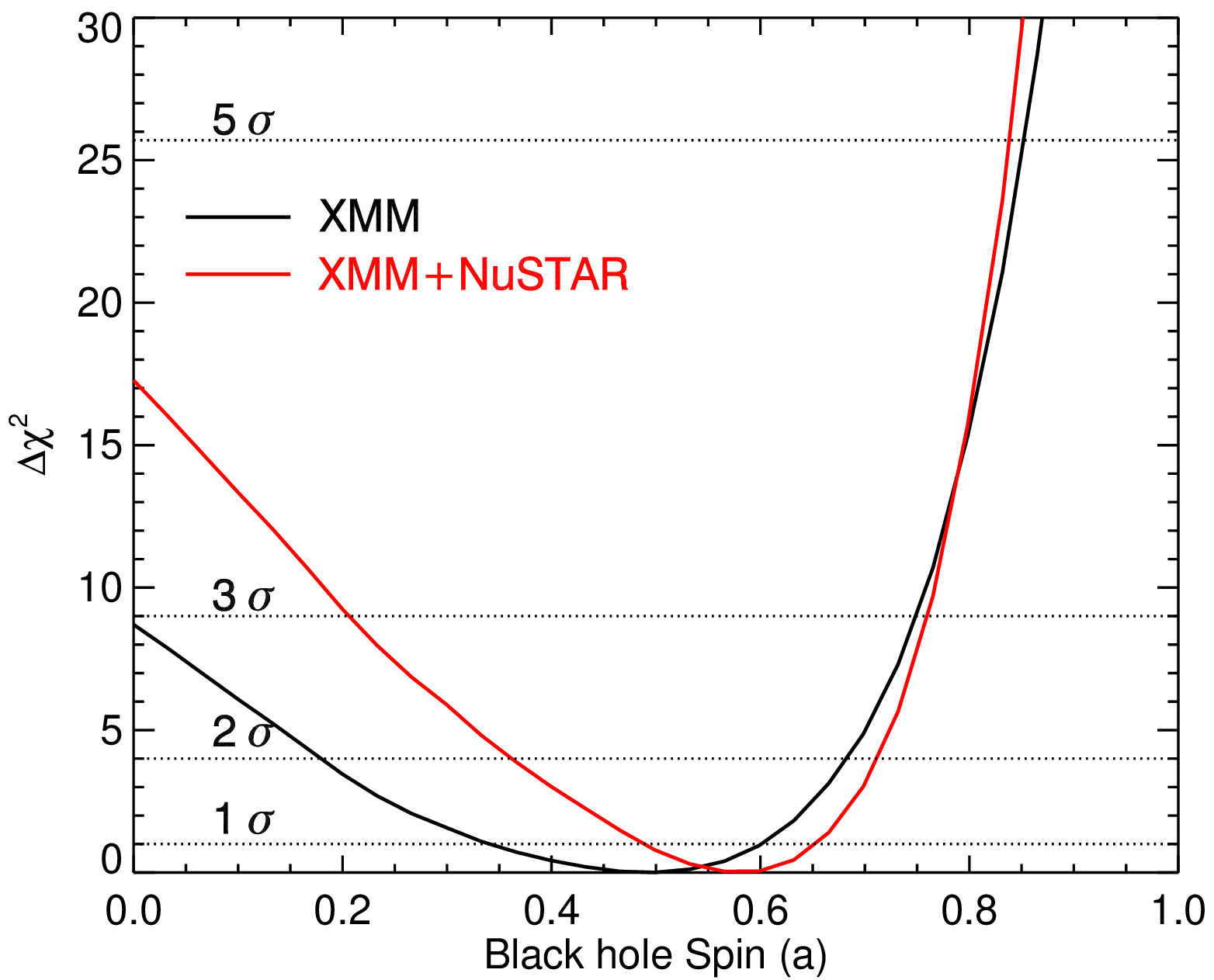, width=\columnwidth}
\caption{\label{spin} Goodness-of-fit variation as a function of the black hole spin. Dashed, horizontal lines indicate $\Delta\chi^2$ values of 1, 4, 9 and 25.7 which correspond approximately to 1, 2, 3 and 5 $\sigma$ confidence levels. }
\end{center}
\end{figure}

We analyzed the {\it NuSTAR}+{\it XMM} spectrum of the NLS1 galaxy SWIFT J2127.4+5654. A relativistic reflection component is found in proximity of an intermediate spin Kerr black hole ($a$=$0.58^{+0.11}_{-0.17}$), confirming past {\it Suzaku} results \citep{mipa09, patrick11} and providing a very accurate measurement of the intermediate black hole spin of this source, thanks to the broad (0.5-80 keV) spectral coverage. Fig. \ref{model} shows the different components of the model, including the primary continuum, the relativistic blurred reflection of the disc and the emission from distant, neutral matter. The contribution from the blurred ionized reflection arising from the accretion disc to the total reflection fraction has been measured and it is described in Sect. \ref{nustaranalysis}. The values obtained are similar to those estimated for other type 1 AGN in \citet{wnf13}. Fig. \ref{spin} shows the goodness-of-fit as a function of the black hole spin: the broader spectral coverage with respect to past analyses allows us to rule out a maximally spinning black hole with a significance $>5\sigma$, and a non-rotating Schwarzschild black hole ($a$=0) is rejected with a significance $>3\sigma$. We find that the accretion disc parameters (disc emissivity and inclination angle) are in agreement with the ones discussed in the past and the steep emissivity ($\epsilon\propto r^{-6.3}$) suggests that the Fe emission arises from the innermost regions of the disc \citep[see][for a more detailed description of the light bending model]{mf04}. Other physical interpretations have been recently discussed for the joint {\it XMM} and {\it Suzaku} analysis of Fairall 9 \citep{lrm12}.
The intermediate black hole spin we measure carries information about the accretion history of the source, and a value of 0.6-0.7 suggests a scenario where the black hole mass in SWIFT J2127.4+5654 is due to a recent black hole merger \citep[see][ for a detailed discussion of the different cases]{volonteri05, bv08} and \citet{vsl07} for a description of the morphology dependent spin evolution.

The high data quality above 10 keV, thanks to the unique {\it NuSTAR} spectral coverage, permits us to measure a cutoff energy E$_{\rm c}$=$108^{+11}_{-10}$ keV with unprecedented accuracy, marginally consistent with prior  {\it INTEGRAL} results where a cutoff energy of $49^{+49}_{-17}$ keV \citep{panessa11} was measured. Using a Comptonization model to reproduce the spectral shape of the primary continuum we find values of kT$_e$=$68^{+37}_{-32}$ keV and $\tau=0.35^{+0.35}_{-0.19}$ with a slab geometry and kT$_e$=$53_{-26}^{+28}$ keV and $\tau=1.35_{-0.67}^{+1.03}$ with a spherical geometry. The power law continuum is an estimate of the power dissipated in the corona and it is $\sim57\%$ of the total 3-80 keV luminosity (${\rm L_{3-80\ keV}^{pow}=}1.92\times10^{43}$ erg s$^{-1}$ for the second orbit, Table \ref{fluxes}). This represents $\sim 7.5\%$ of the bolometric luminosity of the source (${\rm L_{bol}}\sim 2.6\times 10^{44}$ erg s$^{-1}$).

The electron plasma temperatures agree with the ones found in \citet{bmf14} for the Broad Line Seyfert 1 galaxy (BLSy1) IC 4329A, but the lower values for $\tau$ in SWIFT J2127.4+5654 suggest a different geometry in the corona. Indeed, NLS1 galaxies are high rate accreting objects with low black hole masses \citep{mclure04,bz04,nt07}. The lower optical depth might indeed explain why the relativistic reflection in this object ($\sim33\%$ of the total 3-80 keV luminosity) is more prominent than in IC 4329A, where the power law continuum component is higher, being $\sim 87\%$ of the total 5-79 keV luminosity.

SWIFT J2127.4+5654 shows significant lags over a wide range of frequencies.  X-ray time lags are largely agreed to be caused by two separate processes:  by the light travel time between the X-ray source and the ionized accretion disc and by mass accretion rate fluctuations that get propagated inwards on the viscous timescale and cause the soft coronal emission from large radii to respond before the harder coronal emission at smaller radii.  When we measure the lag at a particular frequency, we are measuring the average time lag, and so if both effects are contributing to the lag at a particular frequency, it can be difficult to disentangle which process is responsible for the lag.  In sources with a large soft excess that is dominated by relativistically blurred emission likes (i.e. 1H0707-495 and IRAS~13224-3809), we can easily disentangle these two effects by looking at the lag between $\sim 0.3-1$ keV and $\sim 1-4$ keV.  In these sources, we find that at low frequencies, the hard band lags the soft (due to dominating propagation lags) and at high frequencies, the soft band lags the hard (due to greater reverberation between corona and accretion disc).  For SWIFT J2127.4+5654, the 0.3--1~keV to 1--5~keV frequency-dependent lag shows the typical hard lag of $\sim 1000$~s at frequencies below $3.5\times 10^{-5}$~Hz (left panel of Fig.~\ref{lagfreq}).  It does not show an obvious soft reverberation lag, and this likely because the soft excess is not very strong, due to absorption or because of low iron abundance.  However, we can look for reverberation signatures at higher energies, where there the Fe~K line is strong, and there is little absorption. The frequency-dependent lag in the right panel of Fig.~\ref{lagfreq} shows the 5--8~keV lagging behind the 3--5~keV band up to higher frequencies of $\sim 3 \times 10^{-4}$~Hz.  While some of this lag is likely due to propagation effects (i.e. below $3.5 \times 10^{-5}$~Hz), the high frequency lag ($3.5-30 \times 10^{-5}$~Hz) is likely caused by reverberation, where the emission from the Fe~K line centroid (5--8~keV) lags behind the continuum or red wing of the line (3--5~keV).  The lag-energy spectra support this hypothesis (Fig.~\ref{lagen}).  The low-frequency lag-energy spectrum below $3.5 \times 10^{-5}$~Hz shows a nearly featureless spectrum, increasing with energy (indicative of intrinsic propagation lags in the corona), while the higher frequencies ($\nu=[3.5-30] \times 10^{-5}$~Hz), we see a different shape. There is still little contribution from the soft excess,  but we see a large increase in lag at 5~keV, also where we see the red wing of the Fe~K line.  The lag does not decrease above 8~keV, as is often seen in Fe~K reverberation lags, and this could be an effect of the inclination of the disc \citep{cackett13}, or just too low statistics to confirm a decrease in the lag.

In principle, the amplitude of the lag reflects the light travel time between the primary X-ray source and the accretion disc, and can give an estimate of the size of the corona \citep{wf13}.  However, it is important to take dilution into account before converting the time delay to a physical distance.  For example, at 6.5~keV, the ionised reflection component contributes to $\sim 30$ per cent of the entire variable emission (i.e. not including the cold reflection component), while the continuum contributes to remaining 70 per cent.  Therefore the intrinsic lag will be diluted by 70 per cent, since 70 per cent of the emission comes from the direct continuum. Furthermore, the reference band to which we measure the lag at each energy is also effected by dilution. It is composed of roughly 10 per cent reflection and 90 per cent power law.  If the intrinsic lag in SWIFT J2127.4+5654 is 1000~s, then, including all the effects of dilution, the observed lag at 6.5~keV is $\sim 200s$ (which is close to what we observe).  Given the lower reflection fraction at soft energies below 1~keV, the observed lag should be closer to 50~s, which is also consistent with what we see in the right panel of Fig.~\ref{lagen}.  Assuming an intrinsic lag of 1000~s puts the X-ray source at a height of $\sim 13\ r_{\mathrm{g}}$ above the disc, for a black hole mass of $1.5 \times 10^{7} M_{\odot}$ \citep{maba08}. 

The high-frequency lag-energy spectrum ($\nu=[3.5-30] \times 10^{-5}$~Hz) shows that the Fe~K lag profile is not very broad. Importantly, this timing result is independent of the spectral results, which also shows the emission line is not broad.  All other sources with Fe~K reverberation lags have a maximally spinning black hole, and show Fe~K reverberation lags as low as 3--4~keV. This source only shows a reverberation lag at above 5~keV, as expected for a larger inner radius. This is very strong evidence for an intermediate black hole spin. 

\section{Summary}
We presented a {\it NuSTAR}+{\it XMM} broad band (0.5-80 keV) spectral analysis of the joint observational campaign in 2012 of SWIFT J2127.4+5654. \\
The main results of this paper can be summarized as follows:
\begin{itemize}
\item thanks to the broader spectral coverage with respect to past analyses, a relativistic reflection component is found in proximity of an intermediate spin Kerr black hole ($a$=$0.58^{+0.11}_{-0.17}$), a maximally spinning black hole can be ruled out with a significance $>5\sigma$ and a non-rotating Schwarzschild black hole is rejected with a significance $>3\sigma$;
\item  the high data quality above 10 keV allowed us to measure a cutoff energy E$_{\rm c}$=$108^{+11}_{-10}$ keV with unprecedented precision;
\item the high-frequency lag-energy spectrum shows an Fe~K reverberation lag.  Unlike other maximally spinning black holes that have broad Fe K lags, SWIFT J2127.4+5654 has a narrower lag profile, which independently suggests an intermediate spin black hole.
\end{itemize}
\section*{ACKNOWLEDGEMENTS}
We thank the referee for her/his comments and suggestions that greatly improved the paper.
AM thanks Javier Garcia and Thomas Dauser for the efforts in producing \textsc{xillver} and \textsc{relxill} tables to use in this paper.
AM and GM acknowledge financial support from Italian
Space Agency under grant ASI/INAF I/037/12/0-011/13 and
from the European Union Seventh Framework Programme
(FP7/2007-2013) under grant agreement n.312789. PA and FB acknowledge support from Basal-CATA PFB-06/2007 (FEB), CONICYT-Chile FONDECYT 1101024 (FEB) and Anillo ACT1101 (FEB, PA). MB acknowledges support from the International Fulbright Science
and Technology Award. This work was supported under NASA Contract No. NNG08FD60C, and
made use of data from the {\it NuSTAR} mission, a project led by
the California Institute of Technology, managed by the Jet Propulsion
Laboratory, and funded by the National Aeronautics and Space
Administration. We thank the {\it NuSTAR} Operations, Software and
Calibration teams for support with the execution and analysis of
these observations.  This research has made use of the {\it NuSTAR}
Data Analysis Software (NuSTARDAS) jointly developed by the ASI
Science Data Center (ASDC, Italy) and the California Institute of
Technology (USA).
\bibliographystyle{mn2e}
\bibliography{sbs} 

\end{document}